\documentclass{kapproc}
\usepackage{graphicx}
\usepackage{amsmath}
\usepackage{multirow}
\newlength{\digit}
\newcommand{\tc}[1]{\multicolumn{1}{c}{{#1}}}
\newcommand\tstrut{\rule{0pt}{2.5ex}}

\newlength{\picwidth}
\begin{document}

\articletitle{SPH-based simulation of multi-material asteroid collisions}

\author{T.I.~Maindl$^1$, C.~Sch\"afer$^2$,
 R.~Speith$^3$, \'A.~S\"uli$^{1,4}$, E.~Forg\'acs-Dajka$^{1,4}$, and R.~Dvorak$^1$
}
\bigskip

\affil{$^1$ Institut f\"ur Astronomie, Universit\"at Wien, T\"urkenschanzstra\ss e~17, 
A-1180 Wien, Austria}

\affil{$^2$ Institut f\"ur Astronomie und Astrophysik, Eberhard Karls Universit\"at
T\"ubingen, Auf der Morgenstelle 10, 72076 T\"ubingen, Germany}

\affil{$^3$ Physikalisches Institut, Universit\"at
T\"ubingen, Auf der Morgenstelle 14, 72076 T\"ubingen, Germany}

\affil{$^4$ E\"otv\"os University, Department of Astronomy, 1518 Budapest, Pf.~32, Hungary}

\begin{abstract}
We give a brief introduction to smoothed particle
hydrodynamics methods for continuum mechanics. Specifically, we
present our 3D SPH code to simulate and analyze collisions of asteroids
consisting of two types of material: basaltic rock and ice. We consider
effects like brittle failure, fragmentation, and merging in different
impact scenarios. After validating our code against previously published results
we present first collision results based on measured values for the Weibull flaw distribution parameters of basalt.
\end{abstract}

\begin{keywords}
Minor planets, asteroids -- solar system: formation -- celestial mechanics, stellar dynamics -- methods: numerical -- equation of state -- hydrodynamics
\end{keywords}

\section{Introduction}
\noindent Our goal is to investigate and explain the mechanisms of water delivery processes in early planetary systems. Existing dynamic studies simulate the behavior and collision statistics of asteroid families during and after the Late Heavy Bombardment in the early solar system (e.g., Dvorak et al.\ 2012) and assume a certain water content of the asteroids and complete water delivery to the impact target---usually a (proto)planet. By investigating the impact process itself we will get a more comprehensive description of water delivery starting at an even earlier point in planetary system evolution by answering the question: Under which circumstances does water accumulate in or on bodies resulting from asteroid mergers rather than being lost during the impact?

The impact process can be simulated via a gravitationally bound rubble-pile model where gravitational reaccumulation of fragments governs the formation of (porous) bodies after disruptions (e.g., Richardson et al.\ 2000, 2009, 2012). Another approach is to directly simulate solid state mechanics during an impact. We follow the latter approach using smooth(ed) particle hydrodynamics (SPH) which is a meshless Lagrangian particle
method and was developed by Lucy (1977) and Gingold and Monaghan (1977) for
the simulation of compressible flows in astrophysical context.
For a detailed description of SPH see, e.g., Monaghan (2005) or Sch\"afer et al.\ (2004).
The method has been extended to solid state mechanics by
Libersky \& Petschek (1991). Additionally, a model for the simulation of brittle
failure has been added by Benz \& Asphaug (1994, henceforth referred to as BA94, and
1995). Impacts involving agglomerates such as protoplanetesimals and comets have been successfully simulated using porosity models as described in Sch\"afer, Speith \& Kley (2007), Jutzi, Benz \& Michel (2008) and Jutzi et al.\ (2009).  Also, self gravity has been incorporated successfully.
This makes SPH a promising tool for simulating
planetary and asteroid dynamics (cf.\ Benavidez et al.\ 2012 and
references therein).  While typical grid codes are very well suited to
hydrodynamics in protoplanetary discs they are not so well suited to
treat elasto-plastic behavior or brittle failure which are important
phenomena in collisions of asteroid-like (cf.\ Benz \& Asphaug
1999, henceforth referred to as BA99, Michel, Benz \& Richardson 2004) or moon-sized objects (Jutzi \& Asphaug 2011).

Being a Lagrangian particle method, SPH is suitable for complex
geometries of solid bodies. The continuum of the solid body is
discretized into mass packages which are commonly referenced as
particles. These particles interact by kernel interpolation 
and exchange momentum and energy (cf.\ Sch\"afer 2005). In this first study we introduce the basis of our parallel SPH code and state numerical tests and a first application.

\section{Physical model}
\noindent For modeling the solid bodies we use the Tillotson (1962) equation of state (EOS) as formulated in Melosh (1989). There are two domains depending upon the material energy density $E$. In the case of compressed regions ($\rho \geq \rho_0$) and $E$ lower than the energy of incipient vaporization $E_\mathrm{iv}$ the EOS reads
\begin{align}\label{eq:Pl}
P = \left[ a + \frac{b}{1+E/(E_0 \eta^2)} \right]\rho E + A\mu + B\mu^2
\end{align}
with $\eta = \rho / \rho_0$ and $\mu = \eta-1$. We denote pressure, density, and energy density by $P$, $\rho$, and $E$, respectively. The symbols $\rho_0$, $A$, $B$, $E_0$, $a$, and $b$ are material constants. In case of an expanded state ($E$ greater than the energy of complete vaporization $E_\mathrm{cv}$) the EOS reads
\begin{align}\label{eq:Ph}
P = a\rho E + & \bigg[ \frac{b\rho E}{1+E/(E_0 \eta^2)} 
 + \frac{A\mu}{e^{\beta(\rho_0/\rho-1)}} \bigg] e^{-\alpha(\rho_0/\rho-1)^2}
\end{align}
with two more material parameters $\alpha$ and $\beta$. In the partial vaporization regime $E_\mathrm{iv} < E < E_\mathrm{cv}$, $P$ is linearly interpolated between the pressures obtained via~(\ref{eq:Pl}) and~(\ref{eq:Ph}), respectively. For a more detailed description see Melosh (1989).

The theory of continuum mechanics provides the equations for the
conservation of mass, momentum and energy which describe the dynamics
of a solid body (cf.\ e.g., Sch\"afer, Speith \& Kley 2007).
The conservation of mass is given by the continuity
equation which reads in Lagrangian representation (Einstein notation)
\begin{align*}
\frac{\mathrm{d}\rho}{\mathrm{d}t} + \rho \frac{\partial
v^\alpha}{\partial x^\alpha} = 0.
\end{align*}
The equation for the
conservation of momentum is
\begin{align*} \frac{\mathrm{d}
v^\alpha}{\mathrm{d} t} = \frac{1}{\rho} \frac{\partial \sigma^{\alpha
\beta}}{\partial x^\beta},\quad \sigma^{\alpha \beta} = -P \delta^{\alpha \beta} +
S^{\alpha \beta}
\end{align*}
with the stress tensor $\sigma^{\alpha \beta}$ given by
the pressure $P$ and the deviatoric stress tensor $S^{\alpha \beta}$ and $\delta^{\alpha \beta}$ denoting the
Kronecker delta.
Energy conservation reads
\begin{align*}
\frac{\mathrm{d}E}{\mathrm{d}t} = -\frac{P}{\rho}\frac{\partial v^{\alpha}}{\partial x^\alpha} + \frac{1}{\rho}S^{\alpha \beta}\dot{\epsilon}^{\alpha \beta}
\end{align*}
with the strain rate tensor $\dot{\epsilon}^{\alpha \beta}$ given in~(\ref{eq:repsdot}).

In order to describe the dynamics of a solid body we have to specify the
time evolution of the deviatoric stress tensor $S^{\alpha \beta}$. We
use Hooke's law and define the time evolution as
\begin{align*}
\frac{\mathrm{d}S^{\alpha \beta}}{\mathrm{d}t} = 2 \mu \left(
\dot{\epsilon}^{\alpha \beta} - \frac{1}{3} \delta^{\alpha \beta}
\dot{\epsilon}^{\gamma\gamma} \right) + S^{\alpha \gamma} R^{\gamma \beta} - R^{\alpha \gamma}S^{\gamma \beta}, 
\end{align*}
where $\mu$ denotes the shear modulus. The last two terms involve the rotation rate tensor $R$ and
are rotation terms that are needed since the constitutive
equations have to be independent from the material frame of reference.
We apply the commonly used Jaumann rate form for the rotation terms. The rotation rate and strain rate tensors
are given by
\begin{align}
R^{\alpha \beta} = \frac{1}{2} \left( \frac{\partial v^\alpha}{\partial
x^\beta} - \frac{\partial v^\beta}{\partial x^\alpha} \right), 
\dot{\epsilon}^{\alpha \beta} = \frac{1}{2} \left( \frac{\partial v^\alpha}{\partial
x^\beta} + \frac{\partial v^\beta}{\partial x^\alpha}\right).\label{eq:repsdot}
\end{align}
This set of equations describes the dynamics of an elastic solid body.
In order to model plastic behavior we follow the approach by BA94 and use the von Mises yield criterion where the
deviatoric stress is limited depending on the material yield stress $Y$. We implement this by using a transformed deviatoric stress $S^{\alpha \beta}_\mathrm{vM}$ according to
\begin{align*}
S^{\alpha \beta}_\mathrm{vM} = \min \left[\frac{2\,Y^2}{3\,S^{\alpha \beta}S^{\alpha \beta}},1\right]\cdot S^{\alpha \beta}.
\end{align*}
As basalt is a brittle material we additionally include a damage model
for tensile failure. Physically, fracture is related to the failure of
atomic or molecular bonds. In analogy to the continuum model for solid
bodies a continuum model for fragmentation can be derived (Grady \& Kipp
1980) and has been implemented for these simulations following the
ansatz for SPH by BA94. The distribution $n(\epsilon)$ of flaws activated by a strain level of $\epsilon$ among the SPH particles is given by a Weibull distribution with material parameters $k$ and $m$ according to
$n(\epsilon)=k\epsilon^m$.

\section{Scenarios}
\subsection{Code test}
\noindent To validate our code we use impact simulation results of a lucite bullet impacting a spherical basalt target as published in BA94. We mimic their setup by using the Tillotson equation of state with parameters for basalt and lucite, the material and fracture model constants according to the published values, and only allowing the target to fracture.

\begin{figure*}
\center
\fbox{a) \includegraphics[width=0.18\textwidth, clip]{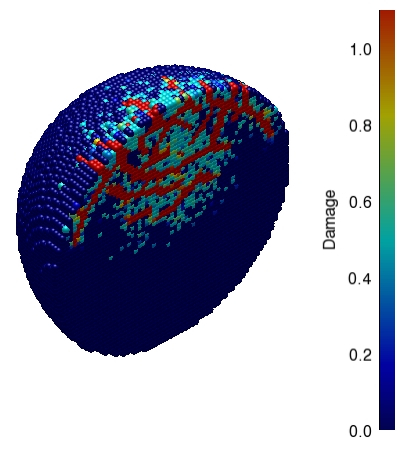}}
\fbox{b) \includegraphics[width=0.18\textwidth, clip]{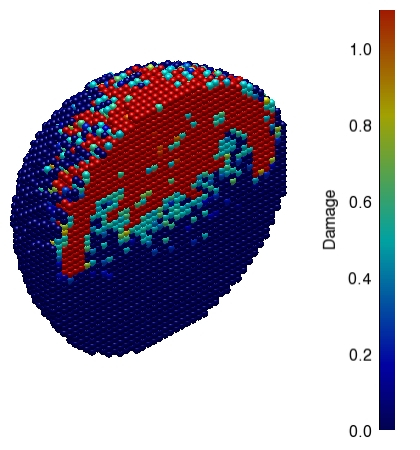}}
\fbox{c) \includegraphics[width=0.18\textwidth, clip]{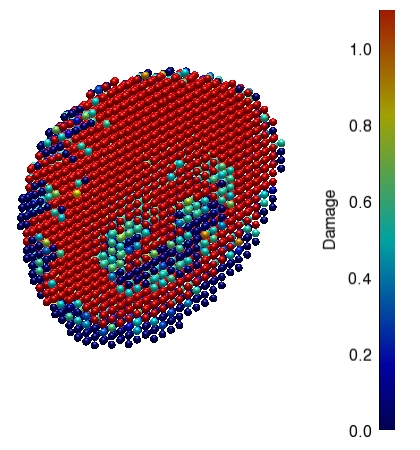}}
\fbox{d) \includegraphics[width=0.18\textwidth, clip]{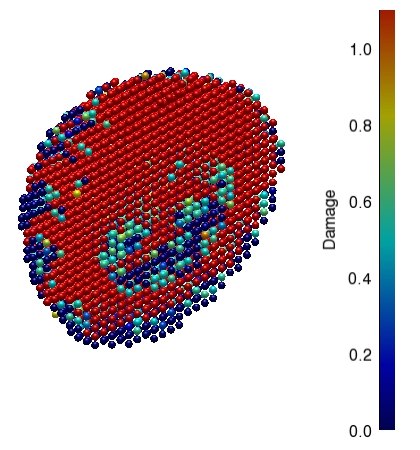}}
\caption{The damage pattern in the test case. The target is split in half to reveal the interior. Scenarios and time after the impact: a) T1, 9.0\,$\mu$s, b) T2, 10.0\,$\mu$s, c) T3, 25\,$\mu$s, d) T3, 40\,$\mu$s.}
\label{f:nak}
\end{figure*}

\begin{table}
\caption{Characteristics of the code tests. The projectile is represented as one SPH particle in all cases.\label{t:BAscen}}
\hfill{}
\begin{tabular}{ccccc}
\hline
\tstrut Scenario   & Number of        & Target radius & Impact velocity  & Impact angle \\
\#         & target particles & (m)              & (m/s)            & ($^\circ$) \\
\hline 
T1         & 145,196   & $3\cdot 10^{-2}$ & $3.2 \cdot 10^3$ & 30\tstrut  \\
T2         & 41,244    & $3\cdot 10^{-2}$ & $3.2 \cdot 10^3$ & 30         \\
T3         & 10,585    & $3\cdot 10^{-2}$ & $3.2 \cdot 10^3$ & 30         \\
\hline
\end{tabular}
\hfill{}
\end{table}
In Tab.~\ref{t:BAscen} we list the simulation characteristics of the code tests. Comparing our results demonstrated in Fig.~\ref{f:nak} with the results of BA94 (cf.\ their Fig.~6) we see a high degree of agreement largely independent from the number of particles used in the simulations. Note the cracks on the surface and the developing subsurface spherical damage patterns as demonstrated in Fig.~\ref{f:nak} a--c which are consistent with Fig.~6 in the reference paper. Also apparent are only minimal changes after about 25\,$\mu$s after the impact (Fig.~\ref{f:nak} c and~d).

\subsection{Impact scenario}
\noindent Our first simulations use a spheroidal target and a smaller, spherical projectile. We allow fracture in both the target and projectile which are composed of different materials.

We treat the collisions from a pure continuum mechanics point of view, i.e., we neglect self gravity as validated for time frames immediately following the impact by BA99.
\begin{figure}
\center
\includegraphics[width=0.55\columnwidth, clip]{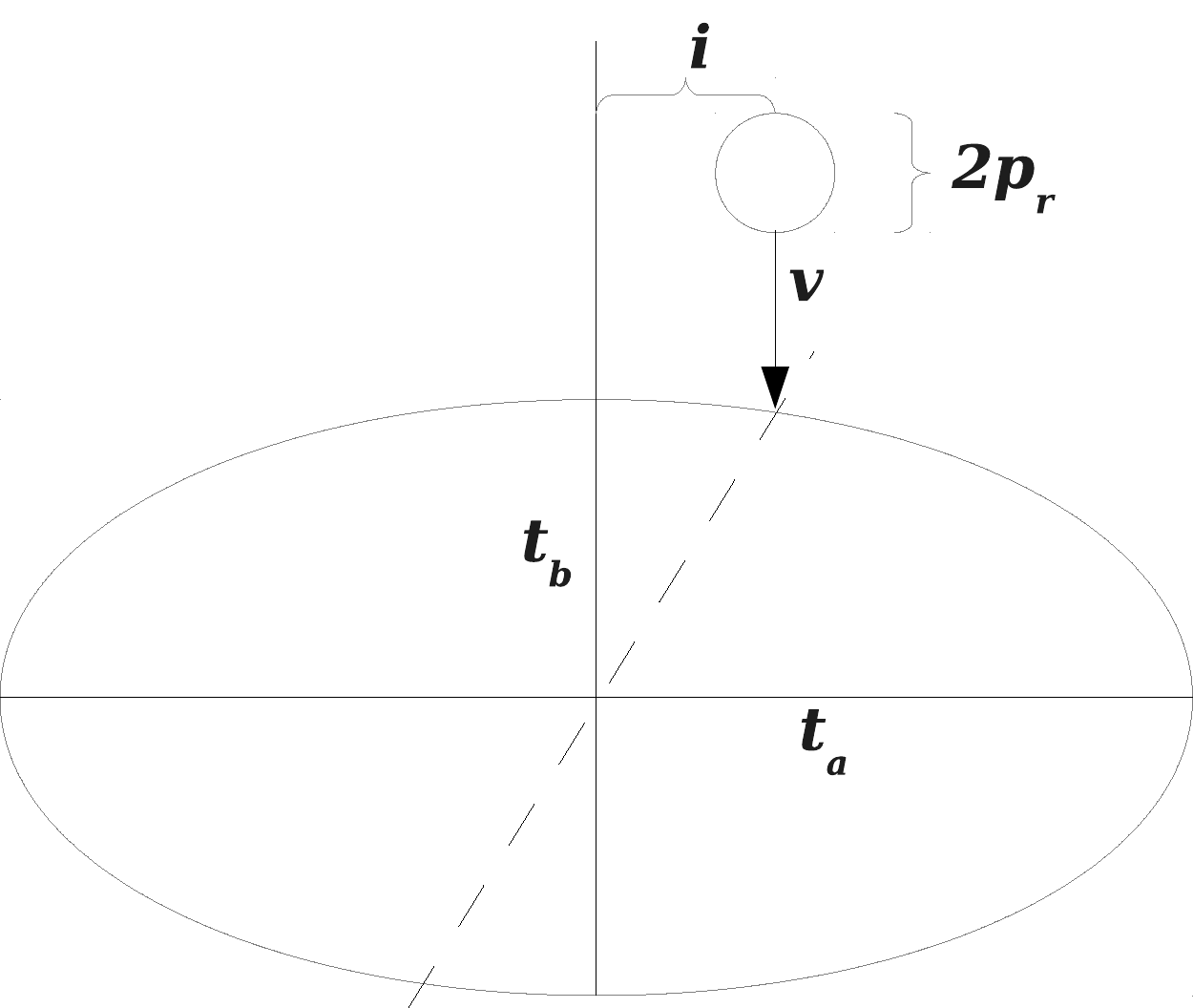}
\caption{The impact geometry (projection onto the $xy$-plane), see text for details.}
\label{f:imp_g}
\end{figure}

As materials for the impacting bodies we use basalt (target) and water ice (projectile). Basalt is a widely adopted material in simulating asteroid collisions using both SPH (cf.\ for example Benavidez et al.\ 2012, BA99) and fluid dynamics codes (cf.\ Korycansky et al.\ 2006).

Figure~\ref{f:imp_g} describes the geometry of the impact scenario: we use a spheroidal target with semi-axes $t_a=10$\,m and $t_b=5$\,m and a spherical projectile (radius $p_r=1$\,m) which impacts the target at a speed of $v=1$\,km/s and an impact parameter $i=3$\,m. The initial particle distribution is chosen such that the projectile and target are composed of 154 and 38,776 SPH particles corresponding to projectile and target masses of about 3.813 and $2.827\cdot 10^3$ tons, respectively.

\begin{table*}
\caption{Tillotson EOS parameters, vaporization energy levels, shear modulus $\mu$, and yield stress $Y$ in SI units, cf.\ Benz~\& Asphaug (1999). Note that $A=B$ is set equal to the bulk modulus.\label{t:EOSpar}}
\hfill{}
\begin{tabular}{l*{10}{r}rl}
\hline
 \tstrut & \tc{$\rho_0$} & \tc{$A$} & \tc{$B$} & \tc{$E_0$} & \tc{$E_\mathrm{iv}$}
         & \tc{$E_\mathrm{cv}$} \\
       & (kg/m$^3$) & (GPa) & (GPa) & (MJ/kg) & (MJ/kg) & (MJ/kg) \\
\hline
Basalt & 2700 & 26.7\hspace{\digit}  & 26.7\hspace{\digit}  & 487 & 4.72\hspace{\digit}  & 18.2\hspace{\digit} \\
Ice    & \hspace{\digit}917 & 9.47 & 9.47 & \hspace{\digit}10 & 0.773 & 3.04 \\
\hline
\end{tabular}
\hfill{}
\vspace*{1ex}

\hfill{}
\begin{tabular}{l*{10}{r}rl}
\hline
 \tstrut & 
         \tc{\multirow{2}{*}{$a$}}
                                & \tc{\multirow{2}{*}{$b$}}
                                & \tc{\multirow{2}{*}{$\alpha$}} 
                                & \tc{\multirow{2}{*}{$\beta$}}
                                & \tc{$\mu$} & \tc{$Y$} \\
       & 
        & & & 
       & (GPa)      & (GPa) \\
\hline
Basalt & 
0.5 & 1.50 &  5.0 & \tstrut 5.0
& 22.7       & \; 3.5 \\
Ice    & 
0.3 & 0.1  & 10.0 & 5.0 & 2.8        & \; 1 \\
\hline
\end{tabular}
\hfill{}
\end{table*}
We use the Tillotson EOS adopting the values given in BA99 following their reasoning on approximating $\rho_0$, setting $A$ equal to the bulk modulus, and $B$=$A$ (see Tab.~\ref{t:EOSpar}). From the same source we use the shear modulus $\mu$ and yield stress $Y$. Young's modulus $E$ is given by $E=9A\mu/(3A+\mu)$.

Similar to the situation with the EOS parameters, it is not easy to get publicly published parameter values for the Weibull distribution governing the density number of flaws. If measurements are not available for the respective material impact experiments can be simulated with varying Weibull parameter values followed by choosing the set of parameters best fitting the experimental results.
As this is not entirely satisfactory we decided to use directly measured
values for basalt $m_\mathrm{basalt}=16$, $k_\mathrm{basalt}=10^{61}\,\mathrm{m}^{-3}$ (Nakamura, Michel~\& Setoh 2007). For ice we adopt the values mentioned in Lange, Ahrens~\& Boslough (1984), $m_\mathrm{ice}=9.1$, $k_\mathrm{ice}=10^{46}\,\mathrm{m}^{-3}$.

Corresponding to the test results we present the evolving damage pattern of our simulations as 3D plots of targets split in half to reveal the inner structure. The cutting plane is aligned with the location of the impact and the origin as sketched by the dashed line in Fig.~\ref{f:imp_g}.

\begin{figure}
\center
\hfill \includegraphics[width=\picwidth, clip]{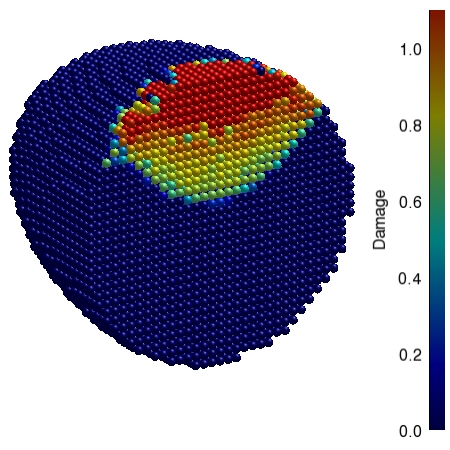}\hfill
\caption{The damage pattern 0.6\,ms after the impact.}
\label{f:ba39k45}
\end{figure}

\begin{figure}
\center
\includegraphics[width=\picwidth, clip]{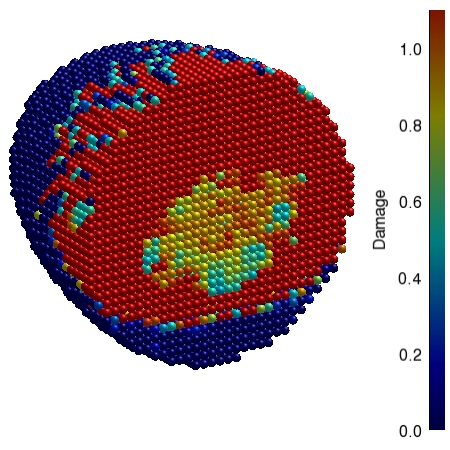}
\caption{The damage pattern 1.5\,ms after the impact.}
\label{f:ba39k54}
\end{figure}

\begin{figure}
\center
\includegraphics[width=0.4\columnwidth, clip]{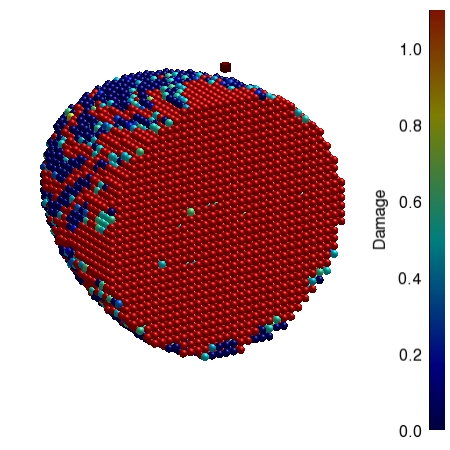}\,
\includegraphics[width=0.4\columnwidth, clip]{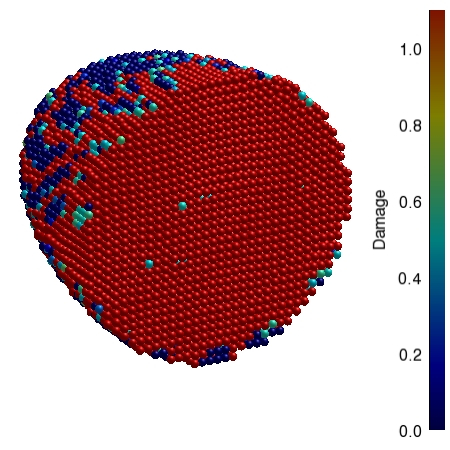}
\caption{The damage patterns 3\,ms (left) and 84\,ms (right) after the impact.}\label{f:ba39klong}
\end{figure}

As can be seen in Figs.~\ref{f:ba39k45} and~\ref{f:ba39k54}, substantial subsurface damage as well as cracks in the surface develop, which is similar to the test scenario. Due to the massive projectile however, the core of the target is also destroyed and only a shell containing undamaged material remains. The surface pattern of undamaged areas intersected by cracks seems to be stable, though as indicated by the comparison of the overall picture 3\,ms and 84\,ms after the impact in Fig.~\ref{f:ba39klong}. The developing inner damage structure will be subject to future higher-resolution simulations.

\section{Conclusions and future research}
\noindent We demonstrated and validated a new SPH code for continuum mechanics with a focus on collisions of bodies in early planetary systems. An immediate application will be connecting to close encounters and collisions as witnessed in n-body simulations (cf.\ S\"uli 2013). From investigating many such encounters we expect to get statistically significant results regarding the merging and fragmentation assumptions in various impact scenarios (velocities, angles, impact parameters, material composition, porosity, body shapes, etc.)---it is more the overall picture than high-accuracy single simulations that are of interest to us at the moment.

Therefore a new, highly parallel code is a necessity to achieve significant results in reasonable computation time. Among others features like XSPH, tensile instability fixes, tensorial correction, artificial viscosity, damage limiting schemes, etc.\ can be adapted for the specific collision scenario under consideration.

\begin{acknowledgments}
\noindent The authors wish to thank \'Akos Bazs\'o for many fruitful discussions.
This paper reflects results from a contribution to the YouResAstro2012 workshop in Budapest, Hungary. We thank the local organizing committee of E\"otv\"os University, Budapest for supporting participation and travel.
This research is produced under a FWF Austrian Science Fund grant (Project ID: S 116-03-N16).
\end{acknowledgments}


\begin{chapthebibliography}{}
  \bibitem{} Benavidez, P.G., Durda, D.D., Enke, B.L., Bottke, W.F., Nesvorn\'y, D., Richardson, D.C., Asphaug, E., Merline, W.J.: 2012, Icar 219, 57
  \bibitem{} Benz, W., Asphaug, E.: 1994, Icar 107, 98
  \bibitem{} Benz, W., Asphaug, E.: 1995, CoPhC 87, 253
  \bibitem{} Benz, W., Asphaug, E.: 1999, Icar 142, 5
  \bibitem{} Dvorak, R., Eggl, S., S\"uli, \'{A}., S\'{a}ndor, Z., Galiazzo, M., Pilat-Lohinger, E.: 2012, AIP Conf.\ Proc.\ 1468, 137
  \bibitem{} Gingold, R.A., Monaghan, J.J.: 1977, MNRAS 181, 375
  \bibitem{} Grady, D.E., Kipp, M.E.: 1980, Int.\ J.\ Rock Mech.\ Min.\ Sci.\ Geomech.\ Abstr.\ 17, 147
  \bibitem{} Jutzi, M., Asphaug, E.: 2011, Natur 476, 69
  \bibitem{} Jutzi, M., Benz, W., Michel, P.: 2008, Icar 198, 242
  \bibitem{} Jutzi, M.,  Michel, P., Hiraoka, K., Nakamura, A.M., Benz, W.: 2009, Icar 201, 802
  \bibitem{} Korycansky, D.G., Harrington, J., Deming, D., Kulick, M.E.: 2006, ApJ 646, 642
  \bibitem{} Lange, M.A., Ahrens, T.J., Boslough, M.B.: 1984, Icar 58, 383
  \bibitem{} {{Libersky}, L.D., {Petschek}, A.G.}: 1991, LNP 395, 248
  \bibitem{} Lucy, L.B.: 1977, AJ 82, 10134
  \bibitem{} Melosh, H.J.: 1989, Impact Cratering: A Geologic Process (New York: Oxford Univ. Press)
  \bibitem{} Michel, P., Benz, W., Richardson, D.C.: 2004, P\&SS 52, 1109
  \bibitem{} Monaghan, J.J.: 2005, RPPh 68, 1703
  \bibitem{} Nakamura, A.M., Michel, P., Setoh, M.: 2007, JGRE 112, E02001
  \bibitem{} Richardson, D.C., Quinn, T., Stadel, J., Lake, G.: 2000, Icar 143, 45
  \bibitem{} Richardson, D.C., Michel, P., Walsh, K.J., Flynn, K.W.: 2009, P\&SS 57, 183
  \bibitem{} Richardson, D.C., Munyan, S.K., Schwartz, S.R., Michel, P.: 2012, LPSC 43.2195
  \bibitem{} Sch\"afer, C.: 2005, Dissertation, Eberhard-Karls-Universit\"at T\"ubingen, Germany
  \bibitem{} Sch\"afer, C., Speith, R., Hipp, M., Kley, W.: 2004, A\&A 418, 325
  \bibitem{} Sch\"afer, C., Speith, R., Kley, W.: 2007, A\&A 470, 733
  \bibitem{} S\"uli, \'A.: 2013, AN, in preparation
  \bibitem{} Tillotson, J.H.: 1962, General Atomic Report GA-3216
  \bibitem{} Weibull, W.A.: 1939, Ingvetensk.\ Akad.\ Handl.\ 151, 5
\end{chapthebibliography}
\end{document}